# Chemical Compositions of Soot Samples in Gold Electrode Capacitors: Molecular Simulations

Vitaly V. Chaban[1] and Nadezhda A. Andreeva[2]

(1) Yerevan State University, Yerevan, 0025, Armenia. E-mail: vvchaban@gmail.com

(2) Peter the Great St. Petersburg Polytechnic University, Saint Petersburg, Russia.

**Abstract**. Electrical breakdown in a dielectric capacitor occurs when the electric field strength across the dielectric material exceeds its breakdown strength. A conductive channel through the dielectric emerges, resulting in a sudden surge of current. Self-healing represents a phenomenon of restoration of a capacitor's performance. The efficiency of self-healing depends on the products of high-temperature decomposition of the electrode and dielectric. We report atomistic simulations of the soot's chemical composition in the case of gold electrodes and four popular dielectric polymers: polypropylene (PP), polyethylene terephthalate (PET), polycarbonate (PC), and polyimide (PI). We unravel that gold atoms form clusters within the carbon-rich soot, limiting their interactions with non-metal elements. The oxygen atoms of PET, PC, and PI act as stabilizers of gold thanks to electrostatic attraction. Compared to zinc, gold equalizes the soot conductivity in all samples but does not impact polymer gasification. The suitability of dielectrics for self-healing must be rated based on the volatile by-products: PP > PC > PET > PI. Mind that PP issues three times more gases than PI. Therefore, the size of semiconducting soot is substantially smaller in the case of PP than in the case of PI. The obtained numerical results provide unprecedented physical insights into the phenomenon of self-healing and clearly drive the efforts to extend the lifespans of the metalized film capacitors.

**Research Highlights**

Semiconducting soot emerges after the electrical breakdown.

Gold atoms form pure gold clusters within the semiconducting soot.

Gold equalizes the electrical conductivities of various soot samples.

Polypropylene is the most efficient polymer for self-healing capacitors.

**TOC Graphic**

The composite simulations reveal the properties of the insulating layer formed after electrical breakdown.

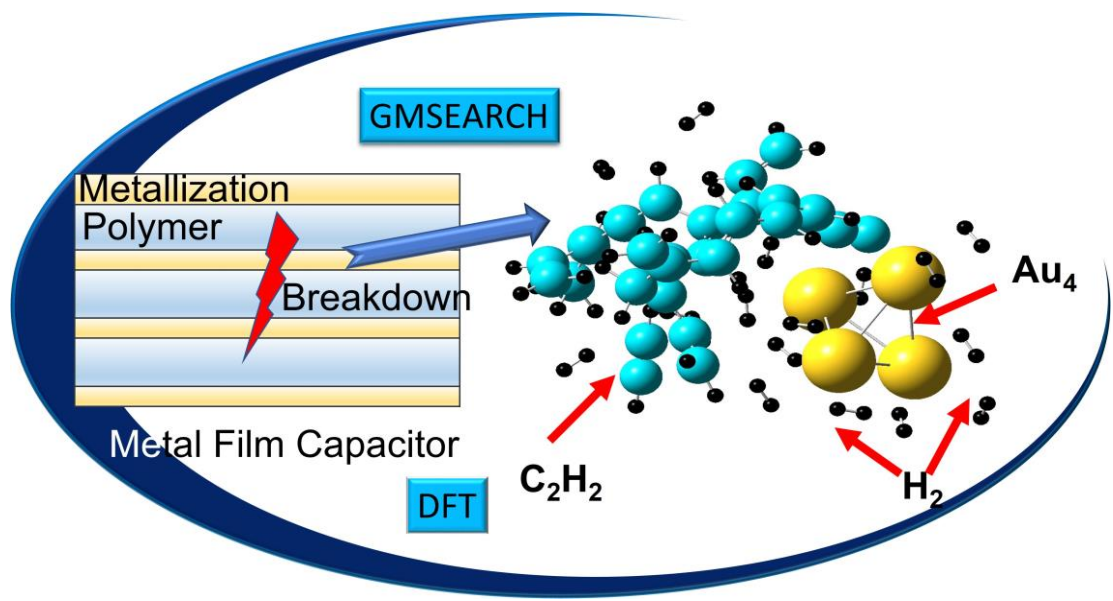

**Introduction**

Dielectric capacitors are essential components in modern electronics, enabling energy storage, filtering, and voltage regulation in various applications.[1-3] However, they are susceptible to electrical breakdown whenever the electric field strength across the dielectric material exceeds its breakdown strength. This phenomenon occurs due to non-uniform structures of insulators and inappropriate electrode/insulator surface contacts. The electrical breakdown can lead to catastrophic failure and compromise circuit functionality. Self-healing mechanisms offer a robust solution to mitigate the effects of breakdown.[3] It allows dielectric capacitors to recover their insulating properties and continue operating in seconds. Nonetheless, some capacity losses happen since a small part of the capacitor decomposes at high temperatures at the time of the breakdown.[4]

An insulator is a dielectric material, which is customarily a carbon-rich polymer, such as polypropylene (PP), polyethylene terephthalate (PET), polycarbonate (PC), or polyimide (PI).[5-8] The PP and PET dielectrics are considered to be the most successful choices for metalized-film capacitors.[9-10] Both of them are applied commercially to dielectric capacitors. In turn, PC and PI exhibit some useful parameters for the functioning of the respective capacitors,[11-13] but their durability is less competitive than that of PP and PET. Researchers tend to link capacitor durability to the efficiency of self-healing after the electrical breakdown.[14]

The chemical products generated during the destruction of the dielectric material at high temperatures play a crucial role in the self-healing process.[14] They influence the dynamics of the insulator's vaporization, arc quenching, and insulation recovery. This in-depth analysis must be performed to explore the intricate relationship between self-healing in dielectric capacitors and the chemical products that emerge upon electrical breakdown. The thermodynamic stabilities, physicochemical properties, and electrical conductivity are of primary importance.[9,13-16]

Gold may be used as an electrode material in dielectric capacitors due to its numerous advantageous properties.[17-18] Firstly, gold exhibits exceptional electrical conductivity, ensuring efficient charge transfer and minimal energy loss within the capacitor. This property is crucial for optimal capacitor performance. It directly impacts the ability of the device to store and discharge electrical energy effectively. Secondly, gold is highly resistant to oxidation and corrosion, contributing to the long-term stability and reliability of the capacitor. Unlike many other metals, gold does not react with oxygen and sulfur (in $H_2S$ gas) over time. Gold is not susceptible to environmental factors, preserving its electrical properties and structural integrity over extended periods.[19] Thirdly, gold has an inert nature, which makes it compatible with a wide range of dielectric materials, allowing for flexibility in capacitor design and optimization for specific applications.[18] This compatibility is essential for tailoring capacitance and breakdown voltage to meet the requirements of diverse electrical circuits and systems. It is important to acknowledge that the cost of gold is a limiting factor in its widespread use, particularly in large-scale or cost-sensitive applications. Zinc and zinc-aluminum alloys are seen to be more sustainable electrode materials in most cases. Despite this limitation, the gold electrode may be considered for use in applications where reliability and long-term stability are paramount factors.[17]

Whereas zinc and aluminum routinely undergo corrosion at elevated temperature and moisture levels, the gold electrodes are chemically inert and, thus, free of the mentioned defects. There have been no academic publications on gold-electrode polymer film capacitors yet. In the present work, we computationally investigate the chemical compositions of the soot that emerges inside the gold-electrode capacitor as a result of the electrical breakdown. The capacitors employing PP, PET, PC, and PI polymers as insulators are considered. The first goal of the work is to determine the chemical form, in which gold is present in the soot. The second goal is to identify the differences in the soot compositions produced out of various dielectric polymers and electrodes out of gold. The third goal is to reveal the role of gold in the electrical properties of the semiconducting soot.

**Methodology**

The kinetic energy injection method was applied to the simulated atomic ensembles belonging to the destroyed gold electrode and various insulator materials in order to reveal a set of stationary points corresponding to these chemical compositions. Each stationary point was uniquely characterized by formation enthalpy and molecular geometry. Plane-wave density functional theory (PWDFT) calculations with pure PBEPBE exchange-correlation functional[20] were used to additionally reoptimize the low-energy geometries and derive the band gaps of the soot samples. In the following, the complete calculation protocols are documented.

The kinetic energy injection method is based on regular perturbations of the system's immediate point (**q**, **p**) in the phase space by adding additional momenta **p**. Next, the excess kinetic energy is gradually deducted via the thermostat, set to room temperature, as the system is propagated upon spontaneous molecular dynamics. The goal of such a perturbation scheme is to allow the system to wander over the entire phase space, irrespective of the activation barriers involved. The atomic forces rely upon the PM7 model Hamiltonian[21-23] to scan the entire potential energy surface as described elsewhere.[24] Note that the Hamiltonian is responsible for the deterministic motion, whereas the perturbation adds the stochastic supplement to the trajectory. The method has been duly tested in dozens of previous works,[25-28] in which low-energy stationary points (global minimum and local minima) had to be located. Herein, the major parameters were set as follows: $10^{-6}$ hartree for the wave function convergence; $3 \times 10^{-2}$ hartree bohr$^{-1}$ for the maximum gradient; $1 \times 10^{-2}$ hartree bohr$^{-1}$ for the root-mean-squared gradient; 5,000 K for the perturbation temperature; 100 fs for the duration of sampling during one cycle; 10 fs and 300 K for the cooling Berendsen thermostat relaxation constant and reference temperature; 0.1 fs for the Verlet time step of atomic nuclei. The elastic boundary conditions were applied to preserve the constant volume of the system in the NVT ensemble. Two hundred non-unique stationary points were obtained for every simulated composition. Note that all enumerated constants used in all employed algorithms correspond to high-temperature simulations and high-energy perturbations. Furthermore, note that the sampling pace of the system is inversely proportional to its size in terms of atomic nuclei. In total, a few million self-consistent-field calculations were performed to derive the reported results.

The PWDFT calculations were conducted using PBEPBE pure density functional theory functional.[29] The van der Waals correction[30] was applied to comprensate for otherwise underestimated dispersion fraction energy of the intermolecular interaction energy. The plane-wave cut-off energy was set to 73 Ry in all systems. The number of k-points for all compositions was set to 27 to sample the Brillouin zone. The cell vectors were optimized along with the optimization of the soot sample geometry.

In-house programs were used to simulate minimum point search.[31-33] Quantum Espresso (version 6.0)[34] was used to optimize the geometries of the periodic systems, optimize the cell vectors, and derive band gaps and electrical conductivities. OpenMopac (https://github.com/openmopac/mopac; version 22) was used to compute PM7 formation enthalpies[21-22,35] whenever needed. The energy landscape exploration was conducted with the in-house software GMSEARCH (version 240203). VMD (version 1.9.3) was used to visualize molecular structures and molecular trajectories.[36]

**Results and Discussion**

Four chemical compositions of the soot forming after the electrical breakdown in a dielectric capacitor were simulated in the present work (Table 1). We assumed the electrode and dielectric to be chemically pure, i.e., no contaminants, soaking liquids, ozonation, and other possible additives were herein considered. The molar ratios between dielectric atoms and electrode atoms were assumed to be approximately 1 to 50. This ratio is an empirically known good practice for commercial capacitor samples. Moreover, it is believed that electrical breakdowns are more destructive for electrodes than for dielectrics because metals exhibit higher thermal conductivities. The reported simulations were undertaken after the complete chemical destruction of the polymers and the gold electrodes. The sizes of the initial systems were chosen in view of the expected computational costs of each single-point calculation and the required number of such calculations to achieve the outlined goals. Note that every system in Table 1 provides 200 non-unique stationary geometries for further analysis.

Table 1. The parameters of the simulated systems, including the chemical compositions, the number of atomic nuclei, and the number of electrons.

| # | Composition | # nuclei | # electrons | Abbreviation |
|---|---|---|---|---|
| 1 | 4 Au + $[C_3H_6]_{10}$ | 94 | 556 | 4 Au + PP |
| 2 | 4 Au + $[C_{10}H_8O_4]_5$ | 114 | 816 | 4 Au + PET |
| 3 | 4 Au + $[C_{22}H_{10}O_5N_2]_2$ | 82 | 708 | 4 Au + PI |
| 4 | 4 Au + $[C_{16}H_{14}O_3]_3$ | 103 | 718 | 4 Au + PC |

Figure 1 depicts the elementary units of the dielectric polymers. Pay attention that the investigated structures are quite different. As well, the chemical compositions differ substantially. Nonetheless, all polymers contain the elements of the second period (carbon, nitrogen, oxygen) and hydrogen atoms to passivate dangling bonds. The major constituent is carbon. It is experimentally known that particularly carbon forms a basis of soot, whereas oxygen and nitrogen may be seen as contaminants. It is customarily mentioned in the field that a carbonized layer emerges as a result of electrical breakdown.

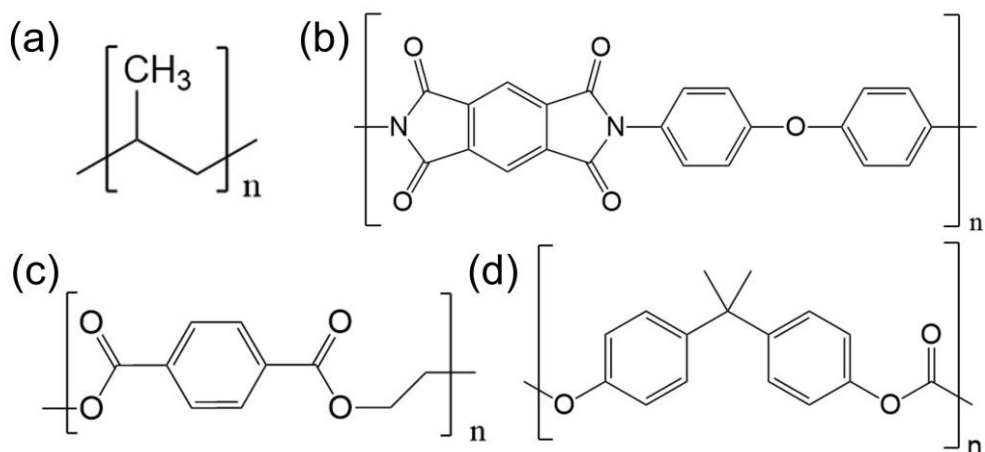

Figure 1. Chemical formulae of the dielectric polymers simulated in the present work are (a) polypropylene, (b) poly-oxydiphenylene-pyromellitimide, better known as a commercial material Kapton, (c) polyethylene terephthalate, and (d) polycarbonate.

The investigation of the potential energy surface assists in understanding the variety of solid and gas by-products obtained out of the available chemical elements in a given proportion. Figure 2 summarizes the standard enthalpies of minimum stationary points detected. To prove the identity of a stationary point, we computed the vibrational frequency profiles at the PM7 level of theory by using the harmonic approximation for atomic vibrations. The minimum state of the soot system must exclude any imaginary frequency by definition.

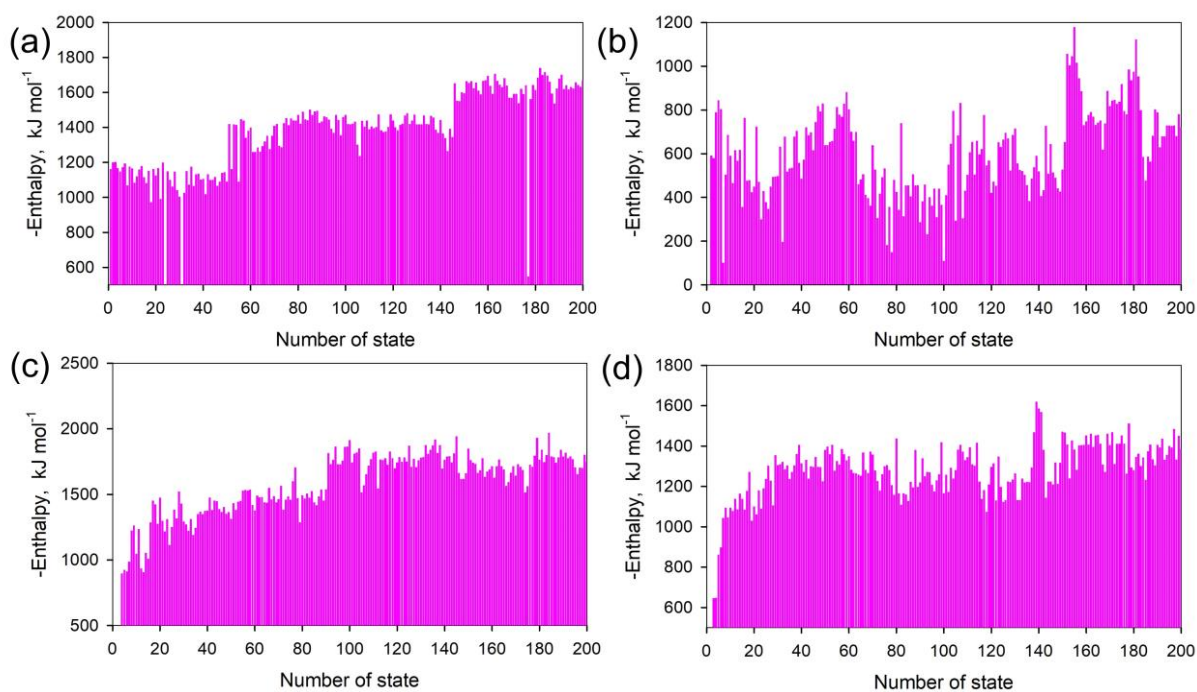

Figure 2. The relative enthalpies of the detected non-unique stationary points (minima) in the simulated systems: (a) 4 Au + PP, (b) 4 Au + PET, (c) 4 Au + PI, and (d) 4 Au + PC. The enthalpy of the highest-energy stationary state is assigned zero for convenience.

We started the potential energy landscape search from highly randomized configurations pertaining to the subplasmic local temperatures in the epicenter of the electrical breakdown. Therefore, the first few stationary points exhibit high formation enthalpies. While these structures possess certain thermodynamic stabilities, they should not be expected in real-world samples. Over macroscopic times, even at room conditions, the system steadily changes them to lower-energy stationary points. After a few perturbation cycles, the geometries of the systems find their relatively stable configurations, which are manifested by the decrease in formation enthalpies by ~1000 kJ/mol. Sporadically, upon potential landscape exploration, additional high-energy configurations of the soot emerge. However, the system swiftly returns to low-energy states afterward. These observations confirm that systems tend to reach more and more stable stationary points during iterations.

The herein simulated chemical compositions are orders of magnitude smaller as compared to the real-world dimensions of the soot samples. Therefore, we must perceive the obtained low-energy configurations as structural patterns, which are repeated in the macroscale samples. The abundance of various patterns is proportional to their formation enthalpies. According to the Boltzmann distribution, the probability of lower-energy pattern occurrence is exponentially higher compared to that of the higher-energy patterns. Our further analysis is primarily based on the most thermodynamically stable atomistic configurations, including the global minimum and a few low-energy local minima exhibiting interesting arrangements of atoms and covalent bonding.

Several factors influence the type and amount of decomposition products generated during electrical breakdown. The chemical composition of the dielectric material is obviously the primary determinant, The decomposition products formed at the subplasmic temperatures are not restricted by any activation energy barriers. Yet, they are restricted by the stoichiometry. For instance, the limited number of oxygen atoms in the closed environment of the capacitors cannot fully oxidize all applicable atoms. Different dielectric materials produce different gaseous by-products, each with its own set of physicochemical properties and effects on self-healing. The energy dissipated during the breakdown event influences the extent of decomposition and the amount of gas generated. The breakdown temperature steadily decreases according to the distance from the epicenter. Higher energy breakdowns lead to more extensive decomposition and a greater variety of chemical products. The temperature and pressure conditions during breakdown affect the chemical reactions involved in the decomposition process, influencing the type and amount of products formed.

Understanding the properties of the decomposition products is the cornerstone for optimizing the regularities of the self-healing process and ensuring long-term capacitor reliability. The thermal conductivity and density of the gas affect the rate of heat dissipation from the arc and the surrounding dielectric material, influencing the arc quenching process and the extent of caused damage. The dielectric strengths of the insulator remainders determine the ability of the insulating zone to withstand the electric field after self-healing, preventing re-ignition of the arc and ensuring device stability. The chemical reactivity of the decomposition products also affects the stability of the capacitor. Some products may potentially react with the dielectric material and the electrodes, leading to unwanted degradation or corrosion. As long as the breakdown products are controlled, the highly precise vision of self-healing can be attained.

The obtained global energy minimum configuration of the soot structure in the 4 Au + PP system (Figure 3) contains 16 $H_2$ molecules and one $C_2H_2$ molecule. The mass fraction of the gas phase is the largest one among all polymers studied. It amounts to 13.9 wt%. Four gold atoms form a tetrahedron. The interatomic distances in the emerged $Au_4$ cluster of gold equal 302, 320, 324, 327, 334, and 339 pm. The graphene-like and other ring structures have not been observed. In turn, the structure of the same composition corresponding to the arbitrarily chosen low-energy local minimum contains 16 $H_2$ molecules and the $Au_4$ cluster. No acetylene molecules emerged in this system, unlike in the global minimum state. The interatomic distances in the gold cluster are 303, 315, 327, 333, 333, and 335 pm. Other low-energy soot structures also tend to include the gold cluster but differ according to the percentage of molecular hydrogen and small hydrocarbons. We conclude that gold atoms, previously belonging to the electrodes, prefer to stay alone after the breakdown.

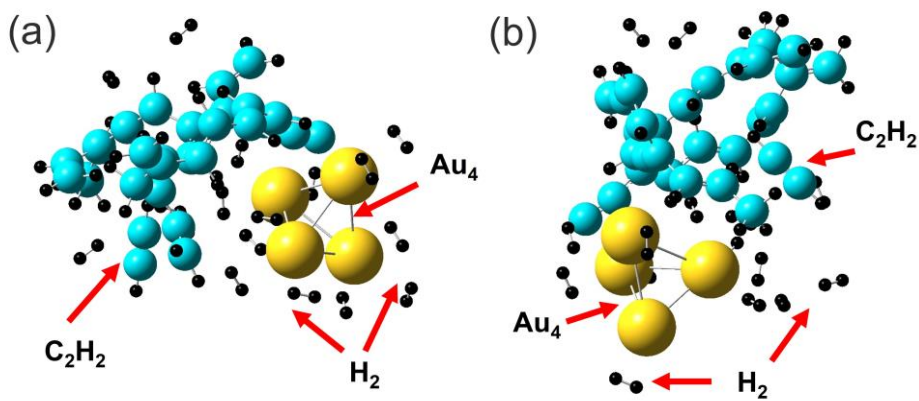

Figure 3. The chemical products formed out of gold electrodes and the PP dielectric :(a) global minimum configuration; (b) local minimum configuration, with a relative enthalpy of +74.6 kJ/mol. The carbon atoms are cyan, the hydrogen atoms are black, and the gold atoms are yellow.

The tendency of Au to self-assemble into a sort of new gold electrode after the electrical breakdown may have various interesting implications. First, the cluster may undermine the efficiency of self-healing by routinely forming a conductive bridge between the electrodes. Second, if the gold cluster attaches to the part of the survived capacitor electrode, this event can signify an outstandingly efficient self-healing. However, the present simulations do not allow one to make further hypotheses. Due to a random outcome, the short-circuiting scenario is deemed to be more realistic if enough gold atoms are available. Whether it so so or not, depends on a particular dielectric capacitor design implemented. Additional modeling efforts using deliberate simulation setups are necessary to understand the extent to which gold clusters can increase their sizes in the presence of the partially or fully destroyed PP dielectric. It may be possible to imagine that two independent soot samples coexist, pure gold soot and carbon-based soot.

The configuration corresponding to the global minimum of energy of the 4 Au + PET system (Figure 4) contains seven $H_2$ and two CO molecules. The weight percentage of these molecules equals 7.3% of the entire system. The soot is composed of various cyclic structures. The gold cluster $Au_3$ emerges as well, whereas one Au atom remains with no covalent bonds. This atom does not form covalent bonds. However, it electrostatically interacts with a few neighboring nucleophilic oxygen atoms. The oxygen atoms polarize the gold atom turning it into the electrophilic interaction center. The scission of the Au cluster does not occur in the PP-containing system but is seen in the PET-containing systems. A logical conclusion from this comparison is the essential role of the oxygen atoms in stabilizing non-clustered gold atoms. Such a conclusion can be validated by considering partial atomic charges of the corresponding Au atoms and oxygen atoms and their interatomic non-covalent distances.

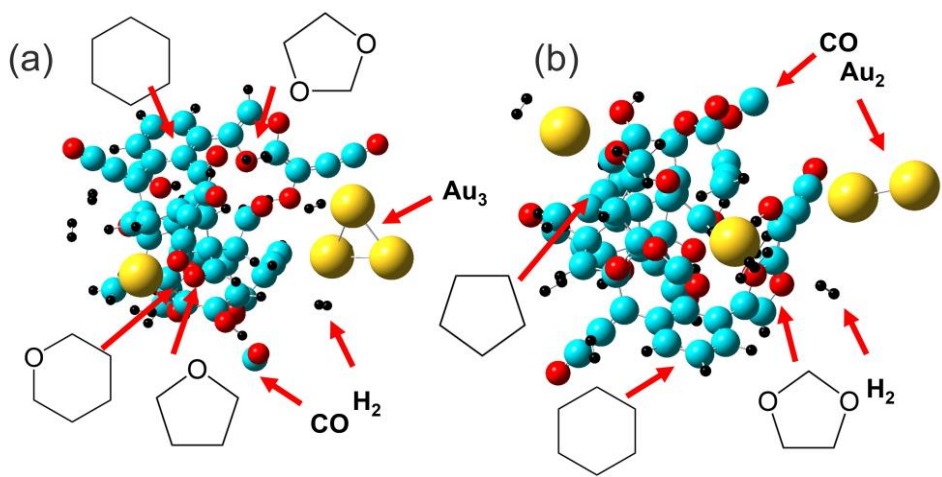

Figure 4. The chemical products formed out of gold electrodes and the PET dielectric: (a) global minimum configuration and (b) local minimum configuration, with a relative enthalpy of +293.1 kJ/mol. The carbon atoms are cyan, the hydrogen atoms are black, the oxygen atoms are red, and the gold atoms are yellow.

For instance, the lone Au atom is located near three oxygen atoms in all detected low-energy configurations of the soot sample. The corresponding Au…O non-covalent distances in the relatively low-energy 106th stationary state are 243, 252, and 355 pm. In turn, the expected Au-O covalent bond length is close to 200 pm. For comparison, the Au-Au covalent bond length amounts to 290 pm. The above estimations have been obtained from the covalent radii of atoms and may insignificantly differ from actual quantum-chemical calculations of the respective chemical bonds and non-covalent interactions.

The partial atomic charges defined according to the Coulson methodology on lone gold atoms are +0.27e, +0.24e, +0.28e, and so on at various low-energy stationary points, 40th, 65th, and 120th, respectively. In turn, the Coulson charges of the three adjacent oxygen atoms amount to -0.53e, -0.53e, and -0.62e. Therefore, the evaluated electrostatic attraction energies binding nucleophilic oxygen atoms and to the electrophilic gold atom are significant. This fraction of non-covalent energy largely compensates for the absence of the lone gold atom in the $Au_4$ cluster.

The distances between gold atoms in its tiny cluster are 298, 309, and 320 pm, each of them corresponding to the length of a plausible Au-Au covalent bond. The low-energy local minimum states are characterized by six $H_2$ molecules and one CO molecule. Gold is present as single atoms and, more rarely, as the $Au_2$ molecule with an interatomic distance of 285 pm. Several cyclic compounds featuring double covalent bonds are present in every revealed stationary point configuration.

The global minimum configuration in the 4 Au + PI system (Figure 5) contains three $H_2$ and one CO molecule. Their mass fraction is 4.5 wt% of the system. We highlight the percentages of the volatile by-products because they are the direct measures of the soot sample size decrease. Small soot samples are unable to conduct electricity being harmless in the context of the capacitor self-healing. The resulting aromatic heterocyclic structures contribute to the increased electrical conductivity of the carbonaceous soot thanks to the conjugated π-π bonds. Gold atoms form small all-metal clusters. The distances between gold atoms are 298, 305, and 313 pm. One gold atom does not form covalent bonds. The local minimum configuration (Figure 5) differs from the global one by 105.4 kJ/mol of energy. It contains the same molecules in the gas phase, but fewer cyclic compounds. The distances between gold atoms in the $Au_3$ cluster are 300, 301, and 322 pm.

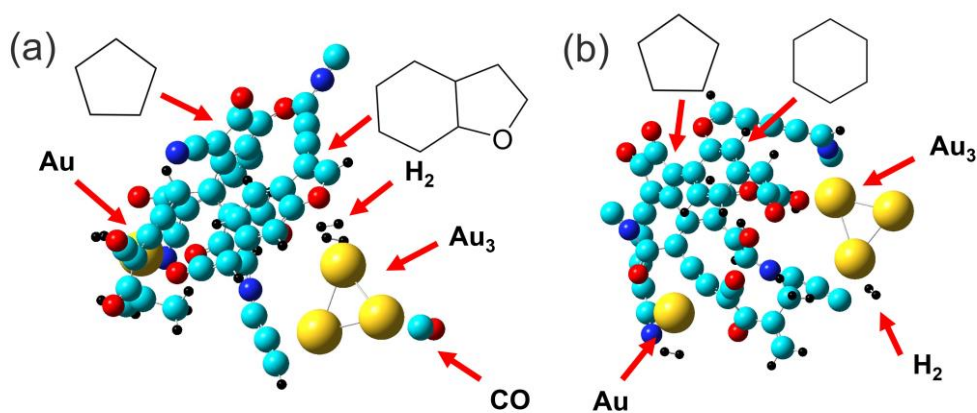

Figure 5. The chemical products formed out of gold electrodes and the PI dielectric: (a) global minimum configuration and (b) local minimum configuration, with a relative enthalpy of +105.4 kJ/mol. The carbon atoms are cyan, the hydrogen atoms are black, the oxygen atoms are red, the nitrogen atoms are blue, and the gold atoms are yellow.

The global minimum configuration of the system composed of 4 Au atoms and PC (Figure 6) includes seven $H_2$ molecules, one CO molecule, and one $CH_4$ molecule, which constitute together 7.6 wt.%. The gold atoms form a cluster-shaped structure, in which they interact via electrostatic forces. This conclusion is derived by the analysis of Au-Au smallest interatomic distances, which amount to 348, 354, 359, and 361 pm. Compare this to the plausible covalent distances for gold of around 300 pm. A rigorous proof would require rigid potential scan simulations or the analysis of the wave function and electron density distribution therein. The obtained structures are thermodynamically stabilized via Au…O attraction patterns discussed above. Ring organic structures out of the polymer atoms do not form. The local minimum (Figure 6) contains the same number of gas molecules. The distances between gold atoms are 333, 356, 363 and 370 pm.

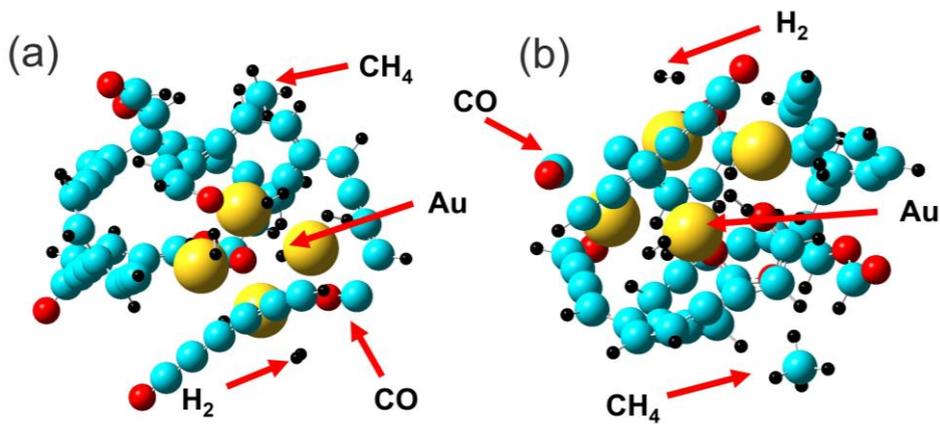

Figure 6. The chemical products formed out of gold electrodes and the PC dielectric: (a) global minimum configuration; (b) local minimum configuration, with a relative enthalpy of +108.2 kJ/mol. The carbon atoms are cyan, the hydrogen atoms are black, the oxygen atoms are red, and the gold atoms are yellow.

As we exemplified above, the electrical breakdown produces both gaseous products and solid products. The solid product, soot, is responsible for the conductivity of the sample, whereas the gaseous products generate high local pressures. The gases are expected to diffuse away from the location of the breakdown and eventually leave the capacitor. Before vanishing, these gases are believed to contribute significantly to the gas pressure buildup. The latter drives the expulsion of molten electrode metal and the formation of the insulating zone between the electrodes. The amount and nature of gas generated depend on the chemical composition of the dielectric material and the conditions of the breakdown. Inert gases can act as arc-quenching agents, promoting the deionization of the plasma and suppressing the propagation of the arc. This helps to minimize the damage to the dielectric material and facilitates faster self-healing. The chemical nature of the solid decomposition products influences the dielectric properties of the insulating zone – carbonaceous soot – formed after self-healing. Some structural patterns contribute to the formation of a thermodynamically stable and high-dielectric-strength insulating layer, whereas others lead to instability over time or unwanted electrical conductivity. For instance, conjugated carbon-based rings with aromaticity are well-known to decrease the HOMO-LUMO band gaps.

Band gap is a convenient electronic property to initially assess the electronic conduction behavior of solid macroscale materials. It is essential to analyze a macroscale sample to get a trustworthy estimate of conductivity. To produce such structures starting from the previously computed low-energy patterns, we employed the plane-wave basis set, which effectively treats the system as periodic and infinite. The periodic boundary conditions work for atomic nuclei and explicitly simulated electrons to represent the replicated system for geometry optimizations and electronic conductivity calculation. In order to proceed with periodic density functional theory calculations, the global minimum configurations, discussed above, were chosen. We hereby assume that all employed methods provide a trustworthy potential energy landscape. Therefore, the low-energy states revealed by non-periodic PM7 are similar, in their major features, to the low-energy states provided by periodic pure density functional theory. The gas molecules were deleted from the low-energy systems prior to simulating electrical conductivities because gases are known not to conduct electricity. The geometries of all systems were optimized again to account for periodic boundary conditions and removed the gas phase. The hereby obtained systems correspond to our theoretical predictions of their identities. Figure 7 depicts periodic unit cells with their right and bottom images.

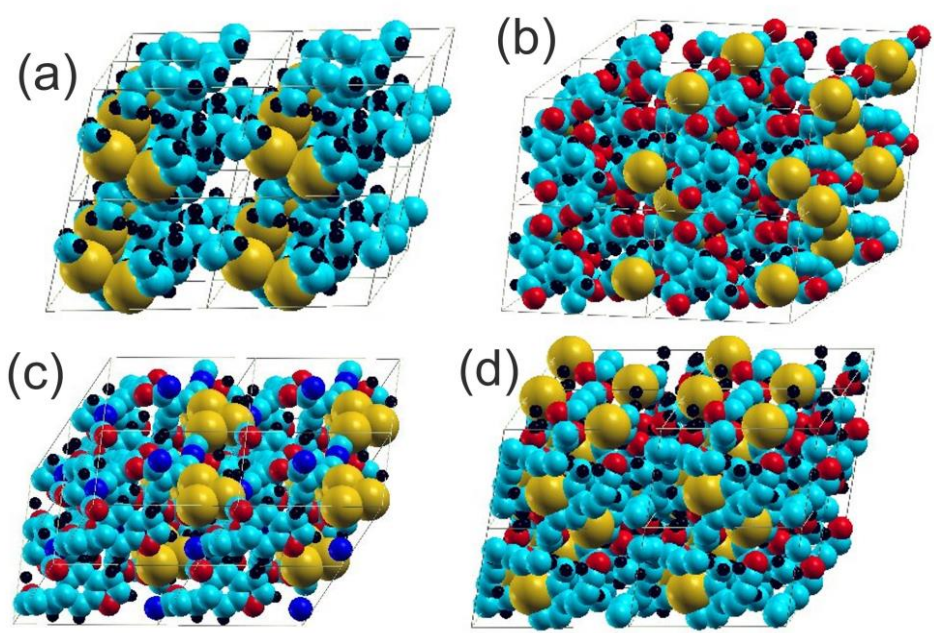

Figure 7. The periodic cells for the DFT-based band gap calculations as obtained from the global minimum molecular configurations of the (a) 4 Au + PP, (b) 4 Au + PET, (c) 4 Au + PI, and (d) 4 Au + PC samples. The carbon atoms are cyan, the hydrogen atoms are black, the oxygen atoms are red, the nitrogen atoms are blue, and the gold atoms are yellow.

Figure 8 depicts the direct band gap for all investigated chemical compositions. The band gap for the "4 Au + PP" system is 0.39 eV, the nad gap for the "4 Au + PET" system is 0.41 eV, the band gap for the "4 Au + PI" system is 0.47 eV, and the band gap for the "4 Au + PC" system is 0.38 eV. As compared to zinc atoms,[37] gold atoms decrease the band gap of a polymer system primarily due to introducing new electronic states, This alteration facilitates charge transfer interactions within the polymer. Gold atoms and their tiny clusters – $Au_2$ to $Au_4$ – introduce localized electronic states within the band gap of the polymer. When these states are near the Fermi level of Au, they effectively reduce the energy difference between the highest occupied molecular orbital and the lowest unoccupied molecular orbital of the destroyed polymer. In this way, the band gap gets effectively lowered.

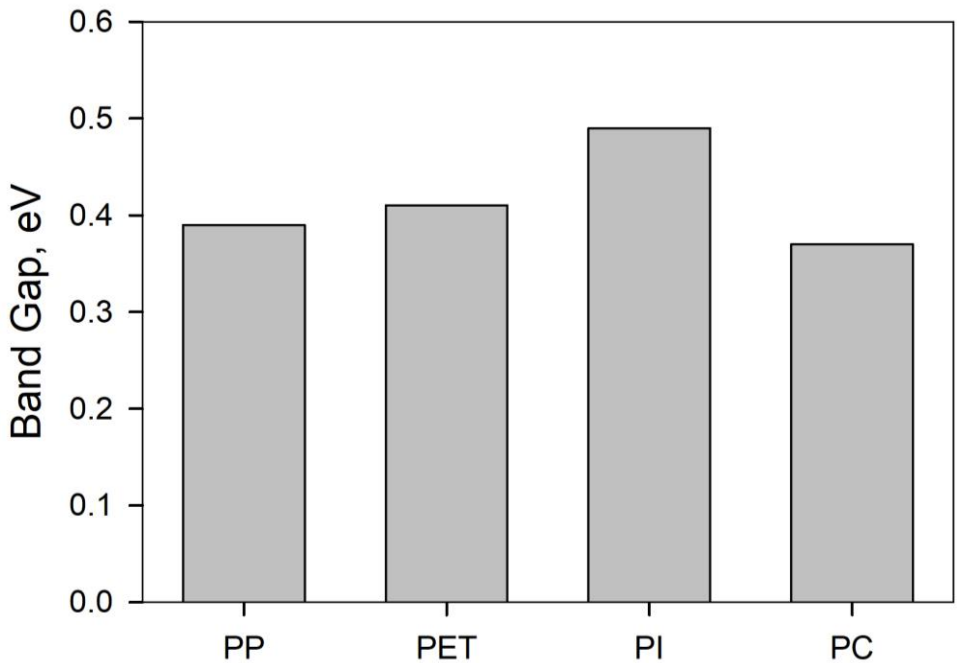

Figure 8. The band gaps of the soot samples obtained out of the Au/PP, Au/PET, Au/ PI, and Au/PC destroyed fragments of dielectric capacitors.

On a related note, Au exhibits a relatively high electron affinity as a heavy metal. It participates in charge transfer interactions with the polymer soot. We exemplified partial atomic charges on the gold atoms above. All gold atoms in the soot possess positive partial charges. Oxygen atoms polarize them in an effort to lower the aggregate potential energy of the simulated. A result of this process is the shifts of the gold atoms' energy levels. As a heavy element, gold possesses valence electrons of relatively high energies, as compared to those of the second-period elements.

When these electrophilic gold atoms are embedded in the inherent structure of the polymer soot as a result of electrical breakdown, they act as electron acceptors. This interaction shifts the energy levels in the former polymer atoms, which are second-period elements. This process reduces the band gap by stabilizing either the valence or destabilizing the conduction band. Gold nanoparticles, in particular, exhibit strong polarizability, which influences the local electronic environment of the polymer. This polarization leads to band bending and also reduces the effective band gap by interacting with the electron cloud of the polymer, making it easier for electrons to transition from valence to conduction bands. The presence of gold increases the degree of conjugation within the emerged carbon-based nanostructures. Gold atoms enhance π-π interactions or facilitate better overlap of molecular orbitals within the semiconducting soot sample, effectively delocalizing electrons over a larger area. This valence electron delocalization reduces the band gap as the electronic structure of the unsaturated carbonaceous structures in the soot becomes more continuous.

Figure 9 represents the localizations of the valence (HOMO) and conduction (LUMO) bands in the "4 Au + PP" system. Note the shape of the unit cell acquired after geometry optimization, including the propagation of the cell vectors. Both the highest orbital of the valence band and the lowest orbital of the conduction band are primarily localized on the gold atoms. Since no ring structures with high-energy delocalized p-electrons exist in this stationary state of this chemical composition, particularly gold clusters are involved in the non-zero electronic conductivity of the corresponding soot sample.

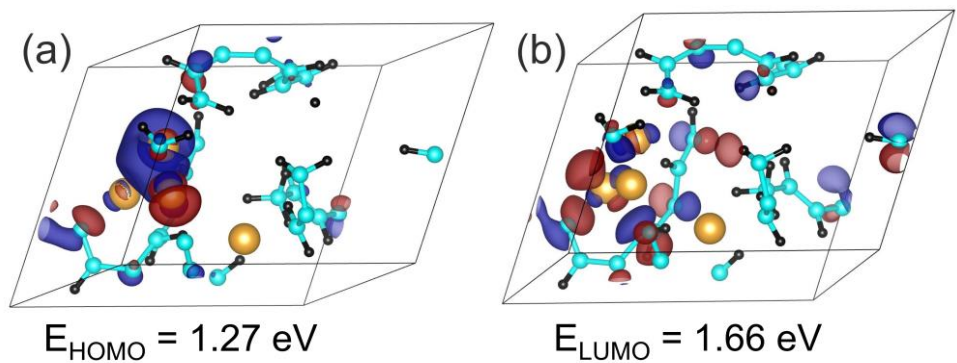

$E_{HOMO}$ = 1.27 eV    $E_{LUMO}$ = 1.66 eV

Figure 9. The spatial localization of HOMO (a) and LUMO (b) in 4 Au + PP system after PWDFT calculations. The carbon atoms are cyan, the hydrogen atoms are black, and the gold atoms are yellow.

Gold atoms may also hybridize with certain molecular orbitals of the polymer, especially if there is a suitable alignment of energy levels. This hybridization creates new hybrid orbitals that bridge the gap between the HOMO and LUMO of the soot. As a result, the energy difference is effectively reduced. Thereby the band gap is decreased and higher conductivity can be expected. Figure 10 summarizes the electrical conductivities of the soot samples: 1.91 kS/m in the "4 Au + PP" system, 2.15 kS/m in the "4 Au + PET" system, 1.80 kS/m in the "4 Au + PI" system, and 1.99 kS/m in the "4 Au + PC" system. We may see that the computed conductivities poorly correlate with the computed band gaps in the same samples. For instance, the PET-containing sample exhibits the second-highest band gap. However, its conductivity appears to be the highest among the compared systems. Compared to the band gap, electronic conductivity is a much more sophisticated property. The position of the Fermi level relative to the band structure determines the material's electrical properties. Furthermore, the number of free electrons or holes available for conduction significantly impacts conductivity. Specific locations of the gold atoms may increase the number of charge carriers and enhance conductivity. The ease with which charge carriers move through the material affects conductivity. In turn, lattice vibrations and irregular structural patterns like defects hinder carrier mobility. While the band provides an estimate of the electric properties of the material, it is essential to compare electronic conductivities to obtain reliable data regarding carbon-based soot conductivity.

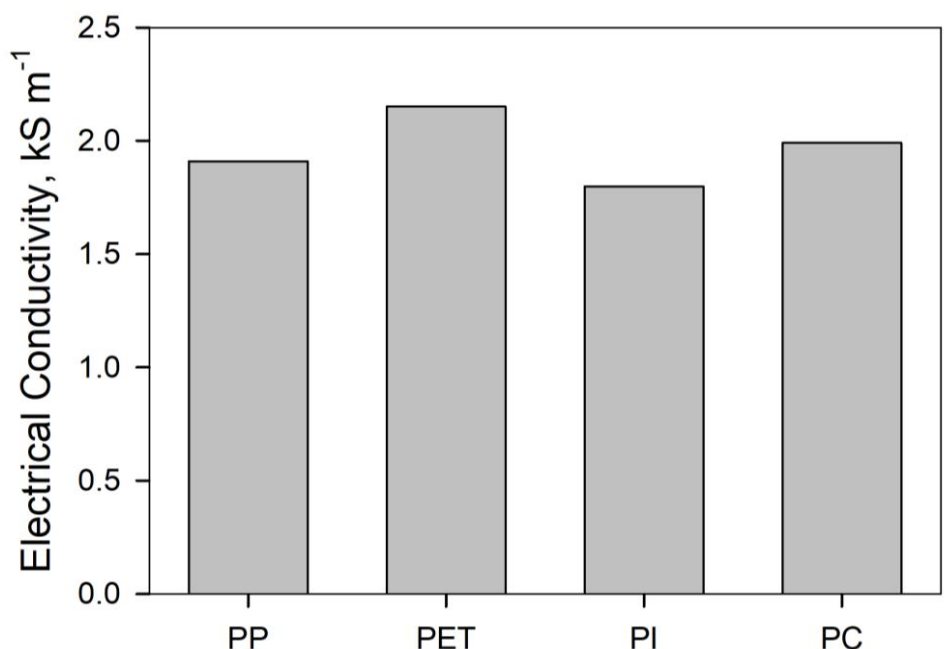

Figure 10. The electrical conductivities of the soot samples obtained out of the Au/PP, Au/PET, Au/ PI, and Au/PC destroyed fragments of dielectric capacitors.

The presence of gold atoms or nanoparticles in a polymer matrix alters the electronic structure of the system through a combination of introducing mid-gap states, enhancing charge transfer, polarizing the polymer matrix, increasing conjugation, and potential orbital hybridization. This results in a reduced band gap, making the material more conductive and altering its optical and electronic properties. The comparison of conductivities to one another reveals that all dielectrics, when destroyed, exhibit very similar conductivities. Compared to zinc electrodes investigated with the same polymers,[37] gold electrodes decrease the band gaps of the soot samples and decrease the conductivities of the soot samples. However, the observed alterations are fairly modest considering the difference between semiconductors and conductors.

All obtained soot samples fall in the range of strong semiconductors with their ~$2\times10^3$ S/m. According to the accepted classification in the field, semiconductors exhibit conductivities of $10^{-6}$ to $10^4$ S/m. Whereas, the conductivity of the true conductors, metals, starts from $10^7$ S/m, i.e., three orders of magnitude higher. The computed soot conductivities fall into the very upper part of the semiconductor range. The elaboration of the methods to decrease these conductivities would help in producing even more reliable dielectric capacitors.

**Verification of the Results**

The hereby discussed results of the research are based on electronic structure simulations involving the latest semiempirical Hamiltonian PM7 and pure density functional theory with a plane-wave basis set. Whereas the first method was used to scan the potential energy landscape, the latter method was used to obtain more accurate geometries of the stationary points of interest and their respective band gaps. Both levels of theory tend to provide realistic structural and thermodynamic properties. The reliability of the electronic properties is less encouraging. Yet, it allows for decent semiquantitative comparison between the systems of similar compositions.

A few analytical techniques may be employed to identify and characterize the decomposition products generated during electrical breakdown: Gas chromatography supplemented by mass spectrometry can be used to separate, identify, and quantify the different gaseous by-products synthesized at high temperatures. Such investigation would provide in-depth insights into the chemical reactions involved in the self-healing process in the case of gold electrodes and routinely used polymers. Fourier transform infrared spectroscopy can be used to identify the functional groups present in the decomposition products, providing information about their chemical structures and potential reactivities. X-ray photoelectron spectroscopy can be used to analyze the elemental composition and chemical states of the decomposition products. This method would provide comprehensive insights into their origin and potential interactions with the capacitor components.

The size of the largest gold cluster $Au_4$ obtained in the present work is obviously limited by the total number of gold atoms in the system. Based on the absence of nucleophilic atoms in PP, we expect gold to form a separate macroscopic phase in the soot. While this phase would exhibit high inherent conductivity, it would likely lack Au atoms to unite electrodes and, hence, cause short-circuit. The precise parameters of this phase would depend on the energy of the specific electrical breakdown and the design of a specific gold-electrode capacitor. This issue and its potential implications call for a dedicated computational and experimental investigation.

**Conclusions and Outlook**

To recapitulate, we reported the computational research of the self-healing process in gold-electrode polymer film capacitors. The chemical products emerging as a result of electrode and dielectric decomposition were characterized. The electrical conductivities of the soot samples were evaluated. The thermally destroyed dielectrics produce significant portions of gases as follows: PP (13.9 wt %) > PC (7.6 wt.%) > PET (7.3 wt.%) > PI (4.5 wt.%). PC marginally outperforms PET. Our results suggest the same trend in polymers as we previously reported for the systems with neglected electrodes, PP > PET > PEN > PPS.[38] Moreover, for the case of the zinc electrodes, the trend is as follows: PP > PC > PET > Kapton.[37] Irrespective of the method of investigation employed and the model Hamiltonian, PP turns out to be the most successful dielectric in the context of self-healing. In turn, high-resistance polymers like PI, Kapton, and PPS support self-healing to a lesser extent. Note that Kapton is a commercial brand of PI, i.e., these two films do not differ in elemental composition.

According to these and previous atomistic simulations, the material of the capacitor electrode does not influence mass fractions of the volatile by-products essentially. Herewith, zinc is much more chemically active and forms covalent bonds with hydrogen, oxygen, and nitrogen atoms. Whereas gold only forms covalent bonds with other gold atoms. In many stationary points, gold does not form covalent bonds, getting stabilized via Coulombic attraction to oxygen atoms, the most nucleophilic interaction centers in the soot samples. PP does not contain oxygen atoms, therefore, gold atoms form pure gold clusters.

Whereas the band gap is generally inversely proportional to the electrical conductivity, allowing for lean and mean assessment of electrical properties, this is not exactly the same in the gold-electrode capacitors. Specifically, according to the band gaps, the conductivities of the soot samples must be as follows: PC > PP > PET > PI. However, the directly computed conductivities are as follows: PET > PC > PP > PI. This observation highlights the importance of deriving electrical conductivities directly while rating the self-healing capabilities of the dielectric polymers.

The electrical conductivities depend on the elemental compositions of the soot samples and their atomistic structures. We see in the present work that conductivities are similar to one another despite a great variety of geometries. This can be explained by a dominant contribution of gold, containing high-energy electrons. These additional energy levels evidently impact the reported values of conductivities. Understanding the intricate relationship between self-healing in dielectric capacitors and the chemical products of electrode and insulator decomposition is critical to thoughtfully improving capacitor technology. By understanding the formation, properties, and impact of these products, we optimize capacitor designs, enhance self-healing capabilities, and maximize capacitor lifespan. The kinetic energy injection method is useful in determining the most probable soot compositions and volatile by-products. In turn, the periodic DFT method handles an infinite model of the soot samples exhibiting informative electronic structures. The opened research avenue promises to boost performance, reliability, and durability, from consumer electronics to high-power industrial systems.

This simulation work confirms the empirical knowledge about dielectric polymers. For instance, PP is routinely referred to as the "gold standard" for metalized-film capacitors. PI, Kapton, and PPS have not led to commercialized capacitors, particularly, due to their inferior longevities. The experimental work[39] evaluates the self-clearing of PP as good, whereas the performances of PET as average.

The field of self-healing capacitor technology has space for upgraded materials and modified dielectric thermal decomposition mechanisms. New dielectric materials that decompose into beneficial gaseous by-products must be probed to foster efficient self-healing and enhanced dielectric capacitor reliability. Techniques are urged to control the decomposition process and selectively generate desired chemical products, optimizing the self-healing process and minimizing unwanted side effects. At the next stage of technological progress in synthetic chemistry, scavenging layers can be incorporated into the capacitor structure to absorb or neutralize harmful decomposition products, preventing long-term degradation and enhancing capacitor lifespan.

**Acknowledgments**

The research was funded by the Ministry of Science and Higher Education of the Russian Federation under the strategic academic leadership program "Priority 2030" (Agreement 075-15-2024-201 dated 06.02.2024). The results of the work were obtained using computational resources of Peter the Great Saint-Petersburg Polytechnic University Supercomputing Center (www.spbstu.ru). We constantly acknowledge Dr. Eugeny Peturhov for his valuable role in installing specific software for our research needs. V.V.C. is an invited research and teaching professor at Yerevan State University (foreign consultant).

**Credit Author Statement**

Author 1: Conceptualization; Methodology Development; Software development; Validation; Formal analysis; Investigation; Resources; Data Curation; Writing - Original Draft; Writing - Review & Editing; Visualization Preparation; Supervision; Project administration; Funding acquisition.

Author 2: Visualization Preparation.

**Conflict of interest**

The authors hereby declare no financial interests and professional connections that might bias the interpretations of the obtained results.